\documentclass[a4paper,11pt]{article}
\pdfoutput=1
\usepackage{jcappub}
\usepackage{array}
\usepackage{amssymb}
\usepackage{amsmath}
\usepackage{mathrsfs}
\usepackage{bm}
\usepackage{graphicx}
\usepackage{hyperref}

\title{ Gravitational waves from resonant amplification of curvature perturbations during inflation}
\author[a,b]{Zhi-Zhang Peng,}
\author[a]{Chengjie Fu,}
\author[c,a]{Jing Liu,}
\author[a,b,c]{Zong-Kuan Guo}
\author[a,b,c]{and Rong-Gen Cai}

\affiliation[a]{CAS Key Laboratory of Theoretical Physics, Institute of Theoretical Physics, Chinese Academy of Sciences, P.O. Box 2735, Beijing 100190, China}

\affiliation[b]{School of Physical Sciences, University of Chinese Academy of Sciences, No.19A Yuquan Road, Beijing 100049, China}

\affiliation[c]{School of Fundamental Physics and Mathematical Sciences, Hangzhou Institute for Advanced Study, University of Chinese Academy of Sciences, Hangzhou 310024, China}

\emailAdd{pengzhizhang@itp.ac.cn}
\emailAdd{fucj@itp.ac.cn}
\emailAdd{liujing@ucas.ac.cn}
\emailAdd{guozk@itp.ac.cn}
\emailAdd{cairg@itp.ac.cn}

\abstract{
	Parametric resonance in a single-field inflationary model with a periodic structure on the potential gives rise to curvature perturbations with large amplitudes on small scales, which could result in observable primordial black holes (PBHs) and concomitant gravitational waves (GWs) induced by curvature perturbations in the radiation-dominated era. 
	In such a model, GWs associated with the PBH formation were investigated in Ref.~\cite{Cai:2019bmk}. 
	In this paper, we consider a stochastic GW background sourced by inflaton perturbations resonantly amplified during inflation. We compute the energy spectra of induced GWs produced both during inflation and in the radiation-dominated era, and find that the peak of the energy spectrum of the former is much higher than that of the latter, but is located at a lower frequency. Moreover, the energy spectrum of induced GWs produced during inflation exhibits a unique oscillating character in the ultraviolet region. Both the stochastic GW backgrounds are expected to be detected by future space-based laser interferometers.
}

\keywords{gravitational waves, inflation, primordial black holes}

\begin{document}
	\maketitle
	
	\section{Introduction}
	The role of primordial black holes (PBHs) \cite{Hawking:1971ei,Carr:1974nx,Carr:1975qj} in astrophysics and cosmology has received considerable attention in recent years. On the one hand, PBHs could account for the LIGO-Virgo gravitational wave (GW) events \cite{Bird:2016dcv,Sasaki:2016jop}, the NANOGrav
	signal \cite{DeLuca:2020agl,Vaskonen:2020lbd,Kohri:2020qqd,Domenech:2020ers,Inomata:2020xad}, and the ultrashort-timescale microlensing events in the OGLE data \cite{Niikura:2019kqi}. On the other hand, the present studies raise the possibility that dark matter comprises of PBHs \cite{Carr:2016drx,Inomata:2017okj,Garcia-Bellido:2017fdg,Kovetz:2017rvv,Carr:2020xqk}.
	The formation of PBHs originates from the gravitational collapse of overdense regions in the early Universe, and such overdense regions may be sourced by
	primordial curvature perturbations with large amplitudes.
	According to current CMB observations, the amplitude of the power spectrum of primordial curvature perturbations on large scales with comoving wavenumbers $k\lesssim 1\;{\rm Mpc}^{-1}$ is constrained to be of order $10^{-9}$ \cite{Akrami:2018odb}. However, on small scales ($k\gtrsim 1\;{\rm Mpc}^{-1}$) the constraints are significantly weaker than those on large scales. 
	If curvature perturbations are enhanced on small scales,
	large-amplitude perturbation modes that enter the Hubble radius in the radiation-dominated era
	can lead to the production of PBHs.
	
	The enhancement of primordial curvature perturbations on small scales can be realized in many inflationary models. A widely studied mechanism is the ultra-slow-roll inflation \cite{Garcia-Bellido:2017mdw,Germani:2017bcs,Motohashi:2017kbs,Ezquiaga:2017fvi,Gong:2017qlj,Ballesteros:2017fsr,Dalianis:2018frf,Gao:2018pvq,Drees:2019xpp,Xu:2019bdp,Fu:2019ttf,Fu:2019vqc,Lin:2020goi,Fu:2020lob,Yi:2020cut}, in which the Hubble slow-roll parameter decreases by several orders of magnitude through flattening the potential or increasing the friction, and so on. Making sound speed close to zero is another feasible way to amplify curvature perturbations \cite{Ballesteros:2018wlw,Kamenshchik:2018sig}. In addition, the parametric resonance of field perturbations, arising from the oscillation of sound speed squared \cite{Cai:2018tuh,Cai:2019jah,Chen:2019zza,Chen:2020uhe} or the inflaton potential \cite{Cai:2019bmk,Zhou:2020kkf,Cai:2021yvq}, can also result in the enhancement of curvature perturbations.
	After their horizon reentry in the radiation-dominated era, enhanced curvature perturbations not only lead to the production of a sizable amount of PBHs, but also inevitably induce a significant GW background since scalar and tensor perturbations couple with each other at the non-linear level (see \cite{Bian:2021ini} for recent comprehensive review about GWs, see \cite{Cai:2018dig}  for discussions about GWs induced by non-Gaussian curvature perturbations, and see \cite{Sa:2007pc}  for studies within loop quantum cosmology).
	
	Unlike some enhancement mechanisms, the exponential growth of scalar field perturbations resulting from the resonance mechanism can source a significant production of GWs during inflation through the second-order effect of perturbations \cite{Cai:2019jah,Zhou:2020kkf}. What's more, it has been noted that the contribution to the total GW background from induced GWs produced during inflation is vitally important. For instance, in Ref. \cite{Cai:2019jah}, the energy spectrum of induced GWs from inflation is comparable to that from the radiation-dominated era, and the total energy spectrum exhibit a unique double-peak pattern. But in \cite{Zhou:2020kkf}, GWs produced during inflation dominate over GWs induced by curvature perturbations in the radiation-dominated era. To sum up, the GW signal from inflation could be an integral part in the profile of the total energy spectrum of GWs for the mechanism of parametric resonance. And then, this specific profile of the energy spectrum of GWs provides us with a chance to distinguish different inflationary models for the generation of large-amplitude primordial curvature perturbations.
	
	In Ref.~\cite{Cai:2019bmk}, however, induced GWs sourced by field perturbations during inflation have not been investigated in the single-field inflationary model with a periodic structure on the potential, which can arise naturally in brane inflation \cite{Bean:2008na} or axion monodromy inflation \cite{McAllister:2008hb,Flauger:2009ab} and give rise to resonance amplification of curvature perturbations.
	To obtain a complete profile of the energy spectrum of induced GWs, which is crucial to test models,
	in the present paper, we shall study the production of GWs during inflation for this model.
	For comparison, we also compute the energy spectrum of GWs induced in the radiation-dominated era.
	The results show that the energy spectrum of GWs produced during inflation has a larger peak value but a lower peak frequency compared with that produced in the radiation-dominated era. Meanwhile, the energy spectrum of GWs produced during inflation exhibits a significant oscillatory behavior in the ultraviolet region. Both contributions to the GW background sum together to provide a unique signal to test theories of inflation.
	
	The organization of the paper is as follows. In Sec. \ref{II}, we review the single-field inflationary model with a small periodic structure on the potential. In Sec.\ref{III}, we present formulae for the energy spectrum of GWs produced both during inflation and in the radiation-dominated era. In Sec.\ref{IV}, we obtain the numerical results. Sec.\ref{V} is devoted to conclusion and discussion. Throughout the paper, we set $c=\hbar=1$, and the reduced Planck mass is defined as $M_{\rm p}=1/\sqrt{8 \pi G}$.

	\section{Model}
	\label{II}
	In this section, we briefly review the enhancement mechanism of primordial curvature perturbations from a single-field inflationary model proposed in \cite{Cai:2019bmk}. In this model, the inflaton potential has the following form,
	\begin{align}
		V(\phi)&=\bar V(\phi)+\xi \cos \left(\frac{\phi}{\phi_\ast}\right) \Theta(\phi-\phi_e)\Theta(\phi_s-\phi),
	\end{align}
	which consists of a simple inflaton potential $\bar V(\phi)$ that is favoured by CMB observation plus a perturbative periodic structure. In the previous and present study, the Starobinsky potential $\bar V(\phi) = \Lambda^4 \left[1-\exp\left(-\sqrt{2/3}\phi/M_{\rm p}\right) \right]^2$ with $\Lambda =0.0032M_{\rm p}$ is considered as a concrete example. The parameter $\xi$ has a dimension of mass to the forth power and determines
	the magnitude of the structure. Due to sub-leading contribution of the $\xi$ term to potential, we have $\xi \ll \Lambda^4$. The parameter $\phi_\ast$ has a dimension of mass and characterizes the period of the structure. $\phi_s$ and $\phi_e$ denote the field values at the start and end points of the perturbative structure respectively, and $\Theta$ is the Heaviside step function.
	
	It is traditional to split the inflaton into a homogeneous background $\phi(t)$ and an inhomogeneous perturbation $\delta\phi(t,{\bf x})$. We focus on the phase when the inflaton rolls from $\phi_s$ to $\phi_e$, and then the equation of motion for the homogeneous part of the inflaton field is given by
	\begin{align}\label{KG}
		\ddot{\phi} + 3H\dot{\phi} + \frac{\partial \bar V}{\partial \phi} - \frac{\xi}{\phi_\ast} \sin \left(\frac{\phi}{\phi_\ast}\right) = 0,
	\end{align}
	with the Hubble parameter $H\equiv \dot a/a$, where $a$ is the scale factor. We work in the spatially flat Friedmann-Robertson-Walker metric, for which the Friedmann equation is given by
	\begin{align}
		3H^2 = \frac{1}{M_{\rm p}^2}\left[ \frac{1}{2}\dot\phi^2 + V(\phi) \right].
	\end{align}
	
	Imposing a condition $|\dot\phi/ (H\phi_\ast)| \gg 1$, Eq. (\ref{KG}) can be divided into
	\begin{align}
		&3H\dot{\phi} + \bar V_{,\phi}\simeq 0\,, \\ &\ddot{\phi}\simeq\frac{\xi}{\phi_\ast}\sin\left(\frac{\phi}{\phi_\ast}\right),
	\end{align}
	which prove to be valid by the numerical calculations. In writing the equation of motion for inflaton perturbations, we will use spatially flat gauge and then the Fourier mode of inflaton perturbations, $\delta\phi_k$, obeys \cite{Hwang:1996gz}
	\begin{align}\label{delta_phi_k}
		\delta\ddot{\phi}_{k}+3 H \delta\dot{\phi}_{k}+\left(\frac{k^2}{a^2}+m^2_{\rm eff}\right) \delta \phi_k=0,
	\end{align}
	with
	\begin{align}
		m^2_{\rm eff}=\frac{\partial^2 \bar V}{\partial\phi^2}-\frac{\xi}{\phi_\ast^2}\cos\left(\frac{\phi}{\phi_\ast}\right)-\frac{1}{M_{\rm p}^2}\left(3 \dot{\phi}^2+\frac{2\dot{\phi}\ddot{\phi}}{H}-\frac{\dot{H}\dot{\phi}^2}{H^2}\right).
	\end{align}
	Introducing a new field variable $\chi_k\equiv a^{3/2}\delta\phi_k$, Eq. \eqref{delta_phi_k} can be rewritten as
	\begin{align}
		\ddot{\chi}_k+\left[\frac{k^2}{a^2}+m^2_{\rm eff}-\frac{9}{4}H^2-\frac{3}{2}\dot{H}\right]\chi_k=0.	
	\end{align}
	We impose another condition $\xi/(H^2\phi_\ast^2)\gg1$, and in the slow-roll regime, the equation becomes \cite{Jin:2020vbl}
	\begin{align}\label{chi_k}
		\ddot{\chi}_k+\left[\frac{k^2}{a^2}-\frac{\xi}{\phi_\ast^2}\cos\left(\frac{\phi}{\phi_\ast}\right)\right]\chi_k=0.	
	\end{align}
	If the field excursion $\Delta\phi\equiv\phi_s-\phi_e$ is small enough so that the evolution of the inflaton background can be simply described as $\phi = \phi_s + \dot \phi_s(t-t_s)$, where any quantity with subscript $s$ denotes it is evaluated at $\phi=\phi_s$, Eq. \eqref{chi_k} can be reexpressed as the well-known Mathieu equation
	\begin{align}\label{Matheiu}
		\frac{d^2\chi_k}{dz^2}+\left[A_k-2q\cos 2z\right] \chi_k=0,
	\end{align}
	with
	\begin{align}
		A_k =\frac{4 \phi_\ast^2k^2}{\dot{\phi}_{s}^2a^2}\,, \qquad q=\frac{2\xi}{\dot{\phi}_{s}^2},
	\end{align}
	after making a change of time variables $2z=\phi_s/\phi_\ast + \dot \phi_s(t-t_s)/\phi_\ast$.
	The existence of an exponential instability, $\chi_k \propto \exp(\mu_kz)$ with a Floquet index $\mu_k > 0$, is a concernful feature of the solutions of the Mathieu equation \cite{Dolgov:1989us,Traschen:1990sw,Kofman:1994rk}.
	This instability corresponds to exponential growth of inflaton perturbations $|\delta\phi_k| \propto a^{-3/2}\exp\left(\mu_k z\right)$ that will result in the amplification of comoving curvature perturbations $\mathcal{R}_k$ ($\equiv H\delta\phi_k / \dot\phi$).
	
	We consider the case of $2\xi<\dot{\phi}_{s}^2$, namely $q<1$, and the most important instability band is located in the region around $A_k \sim 1 \pm q$. For a given $k$ mode, it will be resonantly amplified if $|A_k-1|<q$ is satisfied and the expansion of the Universe will wash out the resonance. Notice that the resonance of $k$ mode occurs when it is deep inside the Hubble horizon since $|\dot\phi/ (H\phi_\ast)| \gg 1$.
	
	\section{GW production}
	\label{III}
	In this section, we present basic formulae for the energy spectrum of GWs sourced by field perturbations during inflation and induced by curvature perturbations in the radiation-dominated era.
	\subsection{In the radiation-dominated era}
	In the conformal Newtonian gauge, we ignore the anisotropic stress and the vector perturbations, and then the perturbed metric has the following form \cite{Ananda:2006af},
	\begin{align}
		ds^2=a(\eta)^2\left\{ -\left(1+2\Psi\right)d\eta^2 +\left[\left(1-2\Psi\right)\delta_{ij}+\frac{1}{2}h_{ij} \right]dx^idx^j \right\},
	\end{align}
	where $\eta$ presents the conformal time, $h_{ij}$ is the second-order transverse-traceless tensor perturbation, and $\Psi$ is the scalar metric perturbation.
	The equation of motion for tensor modes sourced by scalar perturbations reads
	\begin{equation}
		h''_{ij}+2 \mathcal{H} h'_{ij}-\nabla^2h_{ij}=-4 \hat{\mathcal{T}}_{ij}^{~lm}\mathcal{S}_{lm}(\eta,{\bf x}),
	\end{equation}
	where the prime denotes the derivative with respect to the conformal time, $\mathcal{H}=a'/a$ is the conformal Hubble parameter, $\hat{\mathcal{T}}_{ij}^{~lm}$ is the transverse-traceless projection operator, and $S_{ij}$ is the scalar-induced source term.
	The Fourier modes of tensor perturbations are
	\begin{equation}
		h_{ij}(\eta,{\bf x})=\int \frac{d^3k}{(2 \pi)^{3/2}} e^{i{\bf k \cdot x}}\left[h_{\bf k}^+(\eta) e_{ij}^+({\bf k})+h_{\bf k}^\times(\eta) e_{ij}^\times (\bf k)\right],
	\end{equation}
	where $e_{ij}^+({\bf k})$ and $e_{ij}^\times(\bf k)$ are polarization tensors which satisfy $e_{ij}^\lambda({\bf k})e^{\lambda^\prime,ij}({\bf -k})=\delta^{\lambda \lambda^\prime}$, $h_{\bf k}^+(\eta)$ and $h_{\bf k}^\times(\eta)$ denote tensor perturbations with each polarization. In terms of these polarization tensors, the source $\hat{\mathcal{T}}_{ij}^{lm}\mathcal{S}_{lm}(\eta,{\bf x})$ can be expanded as
	\begin{equation}
		\begin{aligned}
			\hat{\mathcal{T}}_{ij}^{~lm} \mathcal{S}_{lm}(\eta,{\bf x})=\sum_{\lambda=+,\times}  \int \frac{d^3k}{(2 \pi)^{3/2}} e^{i{\bf k \cdot x}}
			e_{ij}^{\lambda}({\bf k}) e^{\lambda, lm}({\bf k})  \mathcal{S}_{lm}({\eta,{\bf k}}).
		\end{aligned}
	\end{equation}
	In the following, we omit the polarization index. Now the equation of motion for tensor modes $h_{ij}$ in Fourier space is derived as
	\begin{equation}\label{evolve}
		h''_{\bf k} + 2 \mathcal{H} h'_{\bf k} + k^2 h_{\bf k} = \mathcal{S}(\eta,{\bf k}),
	\end{equation}
	where $\mathcal{S}(\eta,{\bf k}) = -4 e^{lm}({\bf k}) \mathcal{S}_{lm}(\eta,{\bf k})  $. The solution to Eq.\eqref{evolve} can be expressed in terms of the Green function
	\begin{equation}
		h_{{\bf k}}(\eta)=\int^{\eta} d \eta_1 g_{{\bf k}}(\eta, \eta_1) \mathcal{S}(\eta_1,{\bf k}),
	\end{equation}
	where $ g_{{\bf k}}(\eta, \eta_1) $ is the solution of
	\begin{equation}
		g_{{\bf k}} ''+2 \mathcal{H}  g_{{\bf k}} '+k^2
		g_{{\bf k}}=\delta(\eta-\eta_1).
	\end{equation}
	The power spectrum $\mathcal{P}_h$ is formally defined as
	\begin{equation}
		\langle h_{\bf  k}(\eta) h_{\bf p}(\eta) \rangle=\frac{2 \pi^2}{k^3} \delta^3({\bf k}+{\bf p}) \mathcal{P}_h(\eta,{{\bf k}}).
	\end{equation}
	
	In the radiation-dominated era, the source term contributed from scalar metric perturbations at second order can be written as \cite{Ananda:2006af,Baumann:2007zm}
	\begin{equation}
		\mathcal{S}_{ij}(\eta,{\bf x})=4\Psi\partial_i\partial_j\Psi+2\partial_i\Psi\partial_j\Psi-\frac{1}{\mathcal{H}^2}\partial_i(\mathcal{H}\Psi+\Psi^\prime)\partial_j(\mathcal{H}\Psi+\Psi^\prime).
	\end{equation}
	The primordial value of $\Psi_{\bf k}$ is related primordial curvature perturbations as
	\begin{align}
		\langle \Psi_{\bf k} \Psi_{{\bf k}^\prime} \rangle= \frac{2\pi^2}{k^3}\left(\frac{4}{9}\mathcal{P_R}(k)\right)\delta^3({\bf k}+{\bf k}^\prime).
	\end{align}
	The energy spectrum of induced GWs in the radiation-dominated era can be evaluated as \cite{Kohri:2018awv}
	\begin{align}\label{IGW}
		\Omega_{\rm{GW}}&(\eta,k) = \frac{1}{12} \int^\infty_0 dv \int^{|1+v|}_{|1-v|}du \left( \frac{4v^2-(1+v^2-u^2)^2}{4uv}\right)^2\mathcal{P}_\mathcal{R}(ku)\mathcal{P}_\mathcal{R}(kv)\nonumber\\
		&\left( \frac{3}{4u^3v^3}\right)^2 (u^2+v^2-3)^2\nonumber\\
		&\left\{\left[-4uv+(u^2+v^2-3) \ln\left| \frac{3-(u+v)^2}{3-(u-v)^2}\right| \right]^2  + \pi^2(u^2+v^2-3)^2\Theta(v+u-\sqrt{3})\right\}.
	\end{align}
	Taking the thermal history of the Universe into consideration, one can get the GW spectrum at present,
	\begin{equation}
		\Omega_{\rm GW,0}(k)= \Omega_{\gamma,0} \left(\frac{g_{\star,0}}{g_{\star,\rm eq}}\right)^{1/3}  \Omega_{\rm GW}(\eta_{\rm eq},k),
	\end{equation}
	where $\Omega_{\gamma,0}$ is the density parameter of radiation today, $g_{\star,0}$ and $g_{\star,\rm eq}$ are the effective numbers of relativistic degrees of freedom at the present time and at the time $\eta_{\rm eq}$ of the radiation-matter equality, respectively.

	\subsection{During inflation}
	During inflation, since perturbations of the inflaton field are exponentially amplified, they can generate a significant GW background. Now we derive the formulae for the energy spectrum of GWs in the spatially flat gauge.
	One can write down the source term at second order,
	\begin{equation}
		\mathcal{S}_{ij}(\eta,{\bf x})=-\frac{1}{M_{\rm p}^2} \partial_i \delta \phi(\eta,{\bf x}) \partial_j \delta \phi(\eta,{\bf x}).
	\end{equation}
	Then we can obtain
	\begin{equation}
		\mathcal{S}(\eta,{\bf k})=\frac{4}{M_{\rm p}^2} \int \frac{d^3p}{(2\pi)^{3/2}} e^{lm}({\bf k})  p_l p_m \delta\phi_{{\bf p}}(\eta)\delta\phi_{{\bf k-p}}(\eta).
	\end{equation}
	Note that we now take account of induced GWs during inflation which can be approximately described as a de Sitter stage, with $a(\eta)=-1/(H \eta)$ and $\mathcal{H}=a'/a=-1/\eta$. Thus the Green function is given by
	\begin{align}\label{dSGreen}
		g_{{\bm k}}(\eta,\eta_1) = \frac{1}{k^3 \eta_1^2} \Big[ -k(\eta-\eta_1) ~ {\rm cos} k(\eta-\eta_1) + (1+k^2\eta\eta_1){\rm sin} k(\eta-\eta_1)\Big] \Theta(\eta-\eta_1).
	\end{align}
	The power spectrum $\mathcal{P}_h$ of induced GWs from inflation can be derived as
	\begin{align}\label{ph}
		\mathcal{P}_h(\eta_{\rm end},k)&=\frac{2k^3}{\pi^4M_{\rm p}^4}\int_{0}^{\infty}dp p^6\int_{-1}^{1}d{\rm cos}\theta\space{\rm sin}^4\theta \nonumber\\
		&\times\Big|\int_{\eta_s}^{\eta_{\rm end}}d\eta_1g_k(\eta_{\rm end},\eta_1)\delta\phi_p(\eta_1)\delta\phi_{|{\bf k}-{\bf p}|}(\eta_1)
		\Big|^2 ,
	\end{align}
	where $\eta_{\rm end}$ denotes the conformal time at the end of inflation.
	And the present energy spectrum is related to the power spectrum via
	\begin{equation}
		\Omega_{\rm GW,0}(k)\simeq2.70\times10^{-7}\mathcal{P}_h(\eta_{\rm end},k),
	\end{equation}	
	for the frequency $f>10^{-10}\ {\rm Hz}$ \cite{Zhao:2006mm}.
	
	\begin{table}
		\begin{tabular}{>{\centering}p{2cm}>{\centering}p{3cm}>{\centering}p{3cm}>{\centering}p{2cm}>{\centering}p{2cm}}
			\hline
			\hline
			$\textit{Set}$ & $\xi/M_{\rm p}^4$ & $\phi_\ast /M_{\rm p}$ & $\phi_s/M_{\rm p}$ & $\phi_e/M_{\rm p}$  \tabularnewline
			\hline
			
			$1$ & $1.0299\times10^{-14}$ & $1.48\times10^{-4}$ & $4.73$ & $4.72$ \tabularnewline
			
			$2$ & $1.1541\times10^{-14}$ & $2.90\times10^{-4}$ & $4.71$ & $4.66$  \tabularnewline
			\hline
			\hline
		\end{tabular}
		\caption{The parameter sets for producing a sizable amount of PBHs.}
		\label{table1}
	\end{table}

	\section{Numerical results}
	\label{IV}
	In this section, we show the energy spectrum of GWs predicted by this model through numerical calculations.
	We firstly set the e-folding number when the pivot scale $k_p=0.05 \rm Mpc^{-1}$ exits horizon as $N_{\rm CMB}=60$. Then, according to the conditions $|\dot\phi/ (H\phi_\ast)| \gg 1$, $\xi/(H^2\phi_\ast^2)\gg 1$, and $2\xi<\dot{\phi}_{s}^2$ set in the previous analysis, we consider two sets of parameters shown in Table.~\ref{table1}, both of which predict a sizable amount of PBHs with mass around $10^{-12}$ solar mass constituting the most of dark matter.
	
	\begin{figure*}[t]
		\centering
		\includegraphics[width=0.6\textwidth]{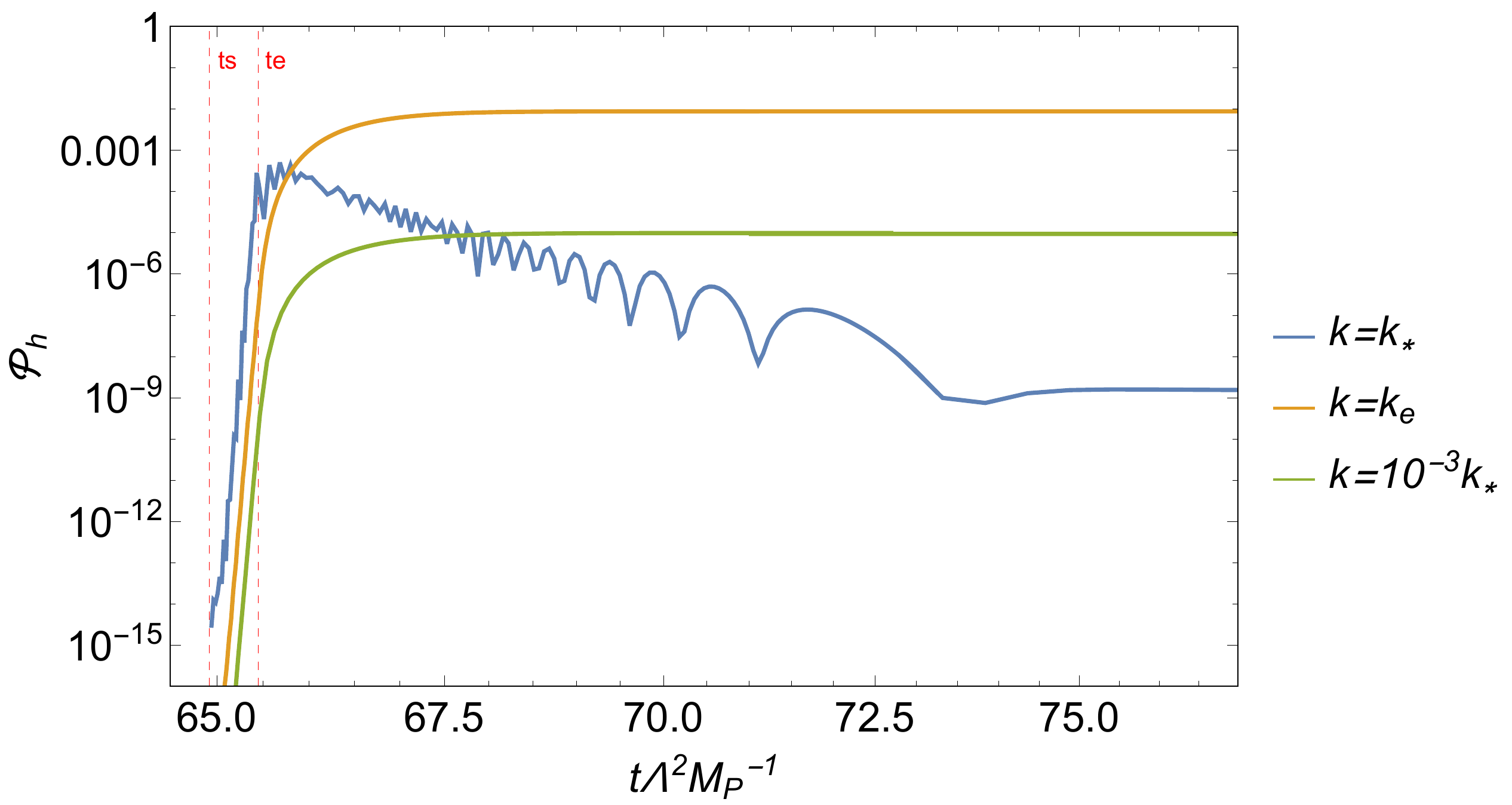}
		\caption{ The time evolution picture of $\mathcal{P}_h$ for $k=k_\ast$, $k=k_e$and $k=10^{-3}k_\ast$ as functions of rescaled time $t\Lambda^2M_{\rm p}^{-1}$.($\xi=1.00\times10^{-14},\phi_\ast=1.48\times10^{-4}$)  }
		\label{fig:phevol}
	\end{figure*}
	\begin{figure*}[t]
		\centering

		\includegraphics[width=0.7\textwidth]{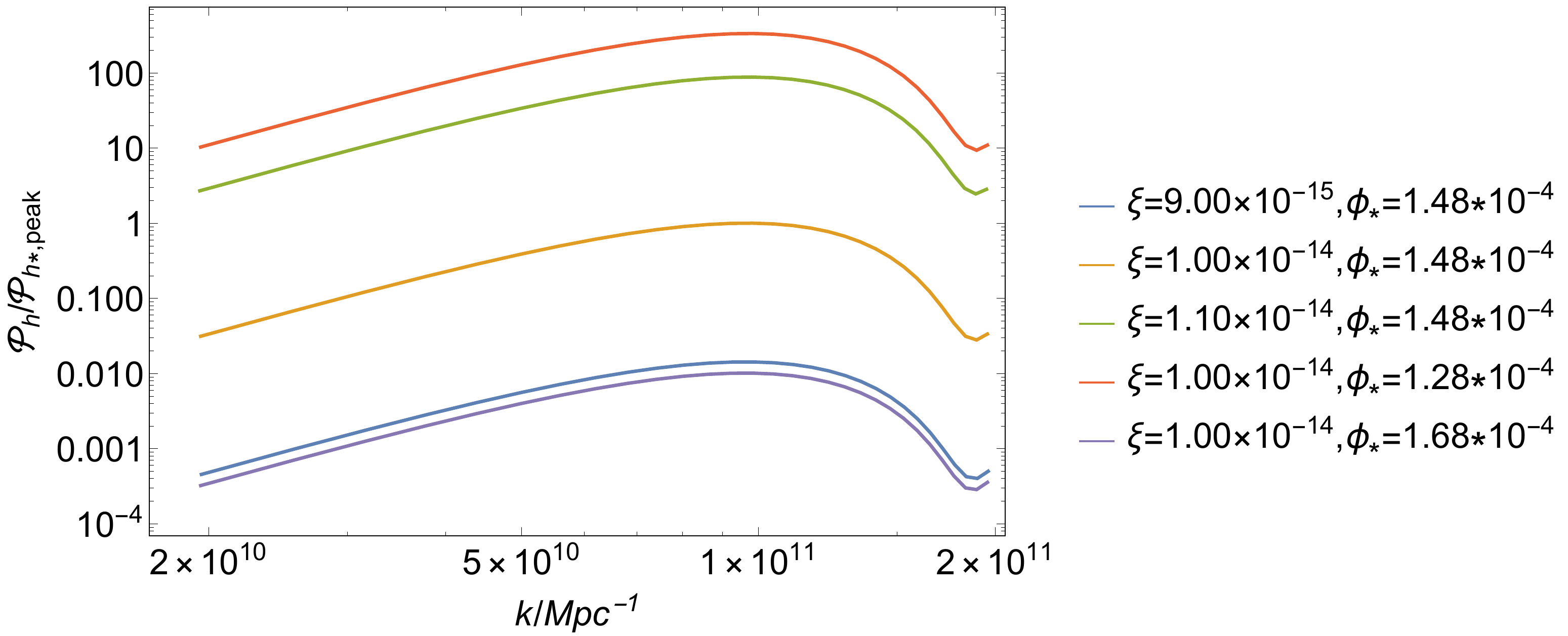}
		\caption{ The inflationary power spectra of induced GWs for different parameter choices. $\mathcal{P}_{h*,\rm peak}$ denotes the peak of $\mathcal{P}_h$ for $ \xi=1.00\times10^{-14},\phi_\ast=1.48\times10^{-4} $. Here we only depict the picture around the peak.}
		\label{fig:phpara}
	\end{figure*}

	We firstly estimate the peak frequency of the power spectrum of GWs during inflation. Taking the power spectrum of primordial curvature perturbations with a narrow peak as an example, according to Eq.~\eqref{ph}, we can find that the main contribution to the integral is only located in the neighborhood of the characteristic scale $k_\ast$ associated with the most enhanced mode of perturbations. In the far-infrared region $k\ll k_\ast$, $\mathcal{P}_h \propto k^3$ since the integral is $k$-independent. In Fig. \ref{fig:phevol}, we depict the time evolution of $\mathcal{P}_{h}$ for $k=k_\ast$, $k=k_e(=a_e H_e)$ and $k=10^{-3}k_\ast$. Note that the modes $k=10^{-3}k_\ast$ and $k=k_e$ leave the Hubble horizon at the end of resonance. From Fig.~\ref{fig:phpara} we can see that both the amplitudes of $\mathcal{P}_{h}$ for $k=k_\ast$ and $k=10^{-3}k_\ast$ are continuously enhanced during the resonance stage due to the exponential growth of the source term, and $\mathcal{P}_{h}$ for $k=k_\ast$ is larger than that for $k=10^{-3}k_\ast$. But after the end of resonance, the source term exponentially decreases with time and soon becomes negligible since $\delta\phi_{\mathbf{k}}\sim a^{-1}$ for $|k\eta|\gg 1$. 
	In the sub-Hubble regime $|k\eta|\gg1$, $h_{ij}$ decays as $a^{-1}$ when the source term is absent, so for $k=k_{*}$, $h_{ij}$ quickly decreases after $t_{e}$. 
	However, for $k=10^{-3}k_{\ast}$, $\mathcal{P}_h$ will quickly get frozen-in on super-Hubble scales. Based on this physical picture, we can estimate that $\mathcal{P}_{h}$ peaks at the mode which leaves the Hubble horizon at the end of the resonance. And the amplitude of GWs with larger $k$ decreases quickly before horizon crossing.
	
	The peak value of power spectrum is difficult to make an analytical estimation because of the complexity of parametric resonance in this model. Specifically speaking, both parameters $\xi$ and $\phi_\ast$ control the rate of amplification, and $\phi_\ast$ also changes the position of characteristic scale $k_{\ast}$. In Fig. \ref{fig:phpara}, we investigate the power spectrum of GWs under two sets of parameters. As expected, we can see that when the period of small structure on the potential remains unchanged, increasing the amplitude of small structure can amplify the peak value of $\mathcal{P}_{h}$. In addition, keeping the amplitude constant, decreasing the period can also enhance the amplitude of $\mathcal{P}_{h}$, because of the increase in the number of oscillations.

	\begin{figure*}[t]
		\centering
		\includegraphics[width=0.4\textwidth]{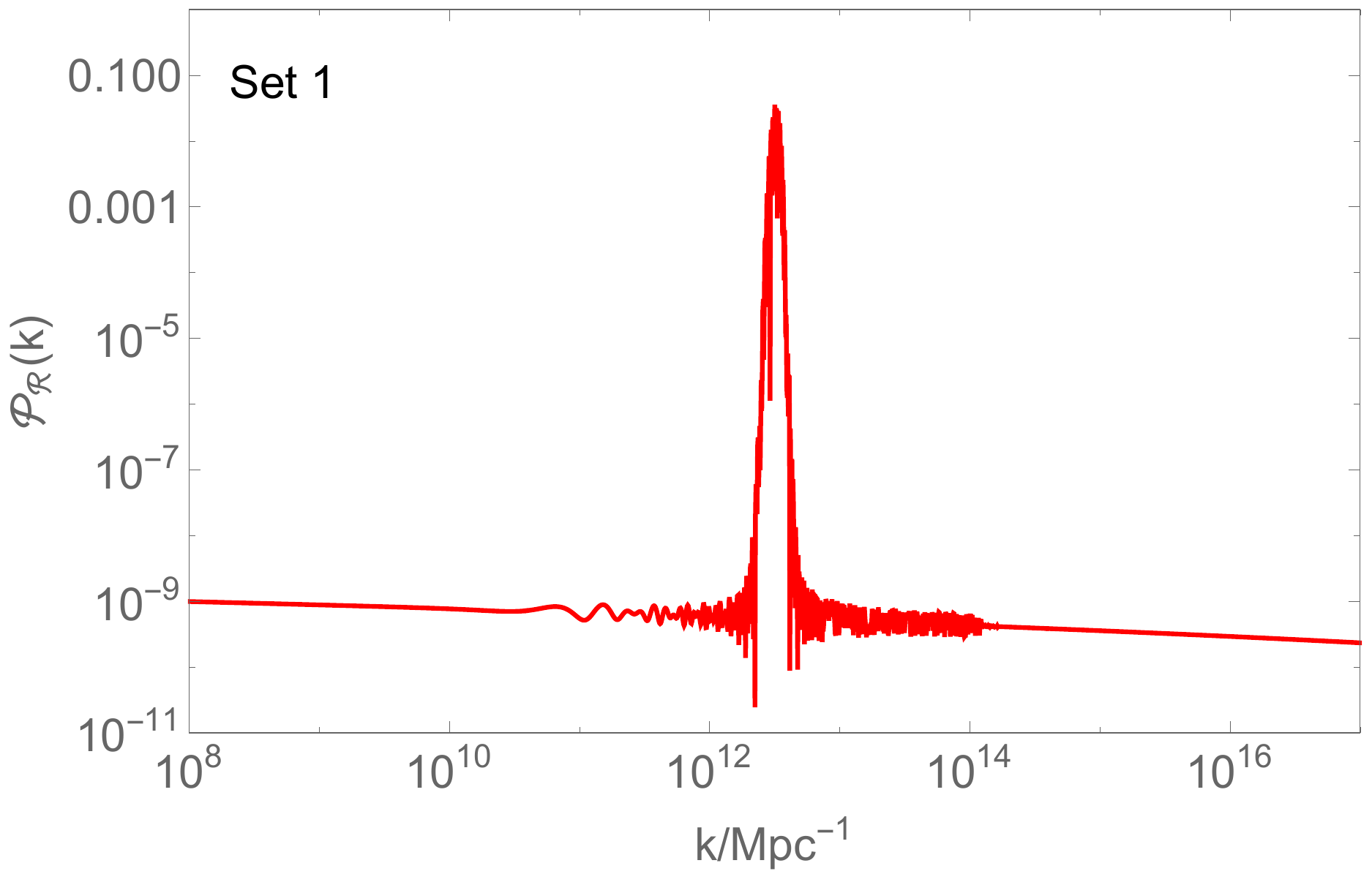}
		\includegraphics[width=0.4\textwidth]{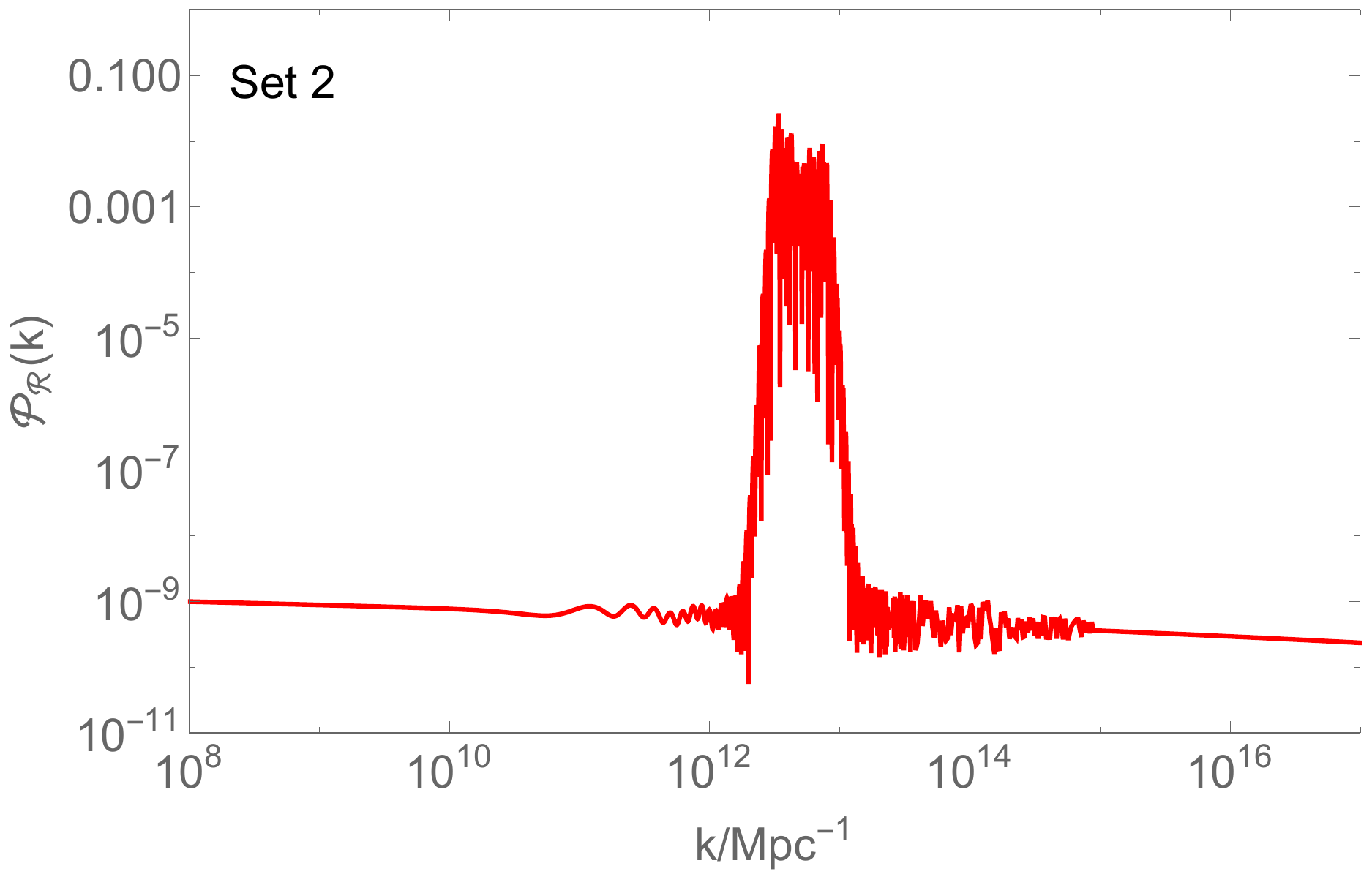}
		\includegraphics[width=0.4\textwidth]{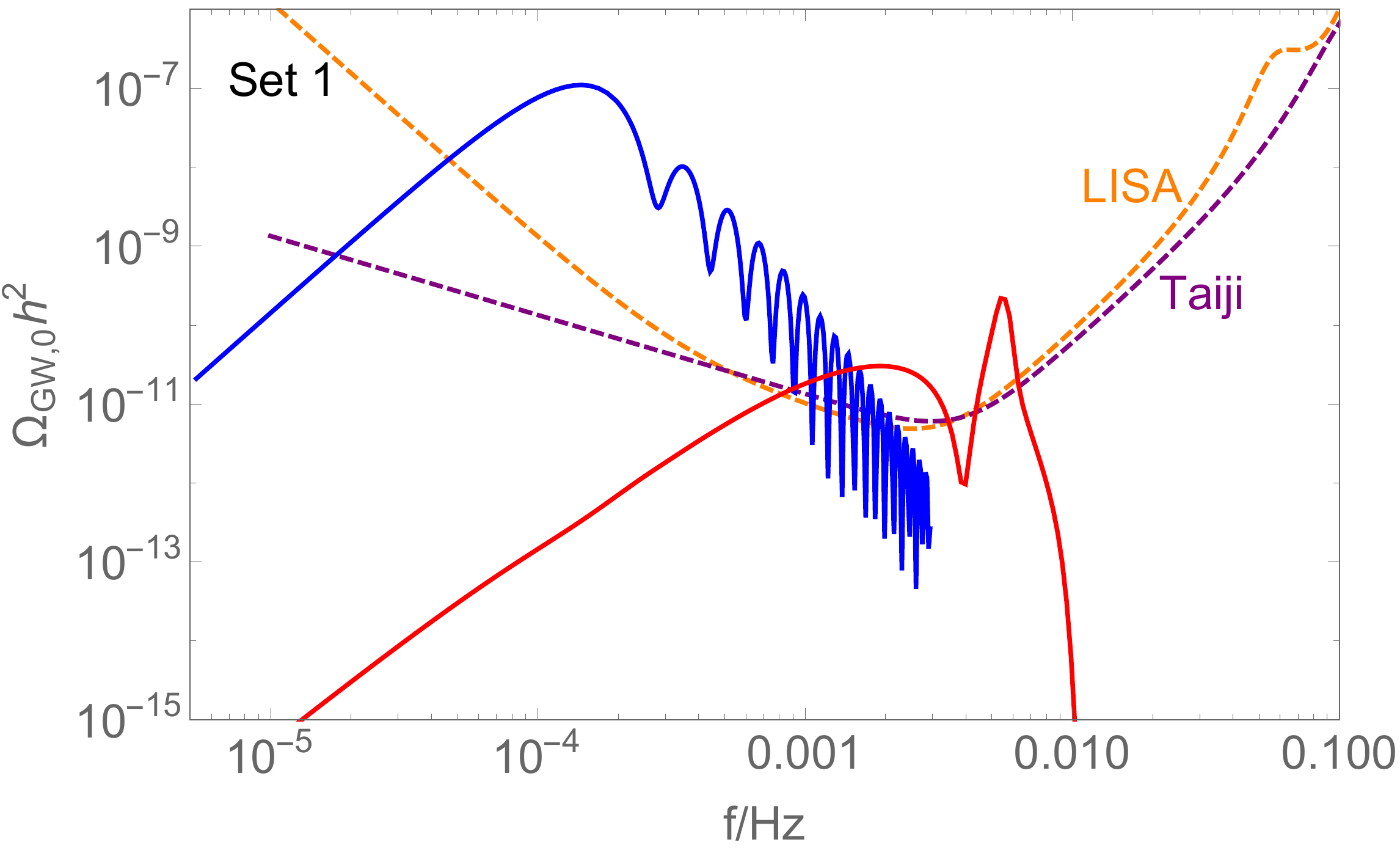}
		\includegraphics[width=0.4\textwidth]{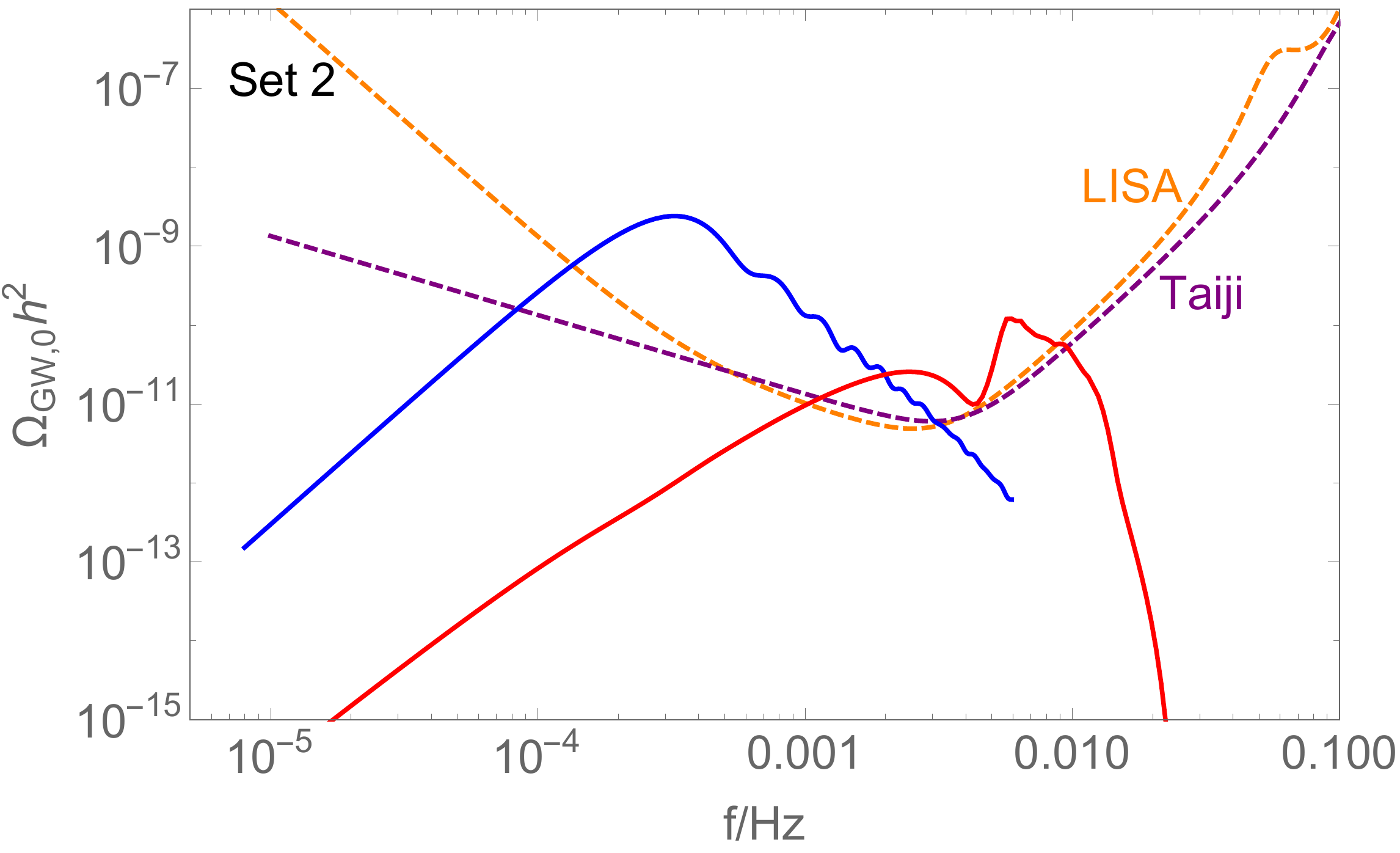}
		\caption{ The resulting power spectra of primordial curvature perturbations in the upper panel and the  present energy spectra of total induced GWs in the lower panel. The blue solid line represents induced GW from inflation and the red solid line represents the one from the radiation-dominated era in the lower panel. The orange and purple dashed lines represent the sensitivity curves of LISA \cite{Audley:2017drz} and Taiji \cite{Guo:2018npi}, respectively. }
		\label{fig:omega}
	\end{figure*}
	
	The upper panel of Fig. \ref{fig:omega} shows the resulting power spectra of primordial curvature perturbations for these two parameter sets, and we can see that the small structure on potential with different profiles results in a narrow peak or a board plateau in the power spectrum.
	This is because the span of resonant modes is proportional to the filed excursion $\Delta\phi$. In the lower panel of Fig. \ref{fig:omega}, we plot the present energy spectrum of total induced GWs and show the sensitivity curves of LISA and Taiji. The peaks of energy spectra of induced GWs, both from inflation and the radiation-dominated era, are above the sensitivity curves of LISA and Taiji. It is interesting to observe that the peak value of the energy spectrum of induced GWs from inflation is much larger than that from the radiation-dominated era. However, the peak frequency of the former is much smaller than that of the latter. This is an important feature to help distinguish these two signals from different eras in future GW detection. Moreover, for the case of narrow spectrum (set $1$), a notable feature is that the energy spectrum of GWs generated during inflation exhibits a unique oscillatory behavior in the ultraviolet region, which is distinct from the one in the radiation-dominated era. But this behavior will gradually disappear as the power spectrum widens. As seen in Fig. \ref{fig:omega}, for the case of broad spectrum (set $2$), the oscillatory structure in the energy spectrum of induced GWs from inflation is too small to be distinguished clearly. 	
	
	\section{Conclusion and discussion}
	\label{V}
	
	In this paper, we have investigated GWs sourced by perturbations of inflaton during inflation in a single-field inflationary model with a small periodic structure on the potential proposed in \cite{Cai:2019bmk}.
	We have calculated and compared the energy spectra of induced GWs during inflation and in the radiation-dominated era, and we compare the differences between them. We find that the energy spectrum of GWs from inflation has a larger peak value and a lower peak frequency compared with that from the radiation-dominated era. Nevertheless, both peaks are located in the sensitive regions of LISA and Taiji. Furthermore, in the case of narrow power spectrum of curvature perturbations, the energy spectrum of induced GWs from inflation displays an oscillatory behavior in the ultraviolet region, which almost disappears in the case of broad spectrum. The detection of such a stochastic GW background, whose energy spectrum includes the features contributed both from inflation and the radiation-dominated era, provides us an inspiring possibility to test the inflationary models. The physical mechanism of the oscillatory structure in the energy spectrum of induced GWs during inflation is still unclear, and we leave this topic to further investigations.

	\section*{Acknowledgments}
	We thank Xing-Yu Yang for useful discussions.
	This work is supported in part by the National Key Research and Development Program of China Grant No. 2020YFC2201501,
	in part by the National Natural Science Foundation of China under Grant No. 12047559, No. 12075297, No. 11690021, and No. 11690022
	and in part by the China Postdoctoral Science Foundation under Grant No.  2020M680689.

	\bibliographystyle{JHEP}
	\bibliography{literature}
	
\end{document}